\documentclass[preprint,showpacs,preprintnumbers,amsmath,amssymb]{revtex4}

\usepackage{graphicx}% Include figure files
\usepackage{dcolumn}% Align table columns on decimal point
\usepackage{bm}% bold math

\begin{document}

% \preprint{APS/123-QED}

\title{Infrasound generation by tornadic supercell storms}% Force line breaks with \\

\author{M. Akhalkatsi}
\affiliation{Centre or Plasma Astrophysics, K.U.Leuven,
Celestijnenlaan 200B, 3001 Leuven, Belgium; Georgian National
Astrophysical Observatory, 2a Kazbegi ave., 0160 Tbilisi, Georgia}

\author{G. Gogoberidze}%
\affiliation{Centre or Plasma Astrophysics, K.U.Leuven,
Celestijnenlaan 200B, 3001 Leuven, Belgium; Georgian National
Astrophysical Observatory, 2a Kazbegi ave., 0160 Tbilisi, Georgia}
% \email{gogober@geo.net.ge}

\date{\today}% It is always \today, today,

\begin{abstract}
Acoustic wave generation by turbulence in the stratified, moist
atmosphere is studied. It is shown that in the saturated moist air
turbulence in addition to the Lighthill's quadrupole and the dipole
sources of sound related to stratification and temperature
fluctuations, there exist monopole sources related to heat and mass
production during the condensation of moisture. We determine
acoustic power of these monopole sources. Performed analysis shows
that the monopole radiation is dominant for typical parameters of
strong convective storms. Obtained results are in good qualitative
agreement with the main observed characteristics of infrasound
radiation by strong convective storms such as total acoustic
power and characteristic frequency.
\end{abstract}

\maketitle

\section{\label{sec:1}Introduction}               % Introduction goes below.

It is long known that strong convective storms, such as supercell
thunderstorms are powerful sources of infrasound \cite{BB71,GH}.
Detailed observations of convective storm generated infrasound
\cite{BB71,NT03} provides that at least two different group of
infrasonic signals could be identified. The first group, with
characteristic period about $1$ s, has been found to be in strong
connection with proto-tornadic structures, funnel clouds and
tornadoes. Based on coincident radar measurements of tornados, which
show strong relationship between funnel diameter and infrasound
frequency, it is usually supposed that these infrasound waves are
generated by radial vibrations of the vortex core \cite{BG00}. The
second group of infrasound signals has the periods from $2$ to $60$
s. Usually the emission appears about 1 hour before observation of
tornado. Although detection of these waves are known to be strongly
correlated with formation of tornado \cite{G88}, it is usually
supposed that these waves are not related with tornado itself and
are caused by convective processes that precede tornado formation.
The acoustic power radiated by convective storm system could be as
high as $10^7$ watts \cite{G88}. Although several reasonable
mechanisms have been suggested to explain this acoustic radiation,
the physical mechanism of the process remains unexplained
\cite{GG75,G88,BG00}.

Broad and smooth spectrum of the observed infrasound radiation
indicates that turbulence is one of the promising sources of the
radiation. Lighthill's acoustic analogy \cite{L52} represents the
basis for understanding of the sound generation by turbulent flows.
Validation of this theory has been shown by various experiments and
numerical simulations (see, e.g., Refs. \cite{WS00,SSBJ01,F03,PSR05} and
references therein). In this approach the flow is assumed to be
known and the sound field is calculated as a small by-product of the
flow. According to this theory in the case of uniform background
thermodynamic parameters interaction of turbulent vortices provides
quadrupole source of sound. The acoustic power of the source was
estimated by Proudman \cite{P52}. But usage of this estimation for
the infrasound radiation from convective storms usually leads to the
underestimation of the acoustic power \cite{GG75,GH,SP07}. It
requires characteristic velocity of the turbulence to be much
greater than it exists in any terrestrial storm system. Recent
analysis of a non-supercell tornado storm simulation performed by
Nicholls, Pielke and Bedard \cite{NPB04} suggests that the
occurrence of the high frequency infrasound coincided with the
development of considerable small-scale turbulence that may have
caused small-scale latent heating fluctuations which appear to be
the main mechanism responsible for generating the infrasound in this
simulation.

In the presented paper we study acoustic radiation from turbulent
convection taking into account effects of stratification,
temperature fluctuations and moisture of the air and using
Lighthill's acoustic analogy. Formulation of the generalized
acoustic analogy \cite{G02} implies: (i) dividing the flow variables
into their mean and fluctuating parts; (ii) subtracting out the
equation for the mean flow; (iii) collecting all the linear terms on
one side of equations and the nonlinear terms on the other side;
(iv) treating the latter terms as the known terms of sound. We show
that in the saturated moist air turbulence in addition to the
Lighthill's quadrupole and known dipole sources of sound related to
stratification and temperature fluctuations, there exist monopole
sources related to heat and mass production during the condensation
of moisture. It appears that infrasound radiation from convective
storms should be dominated by acoustic source related to the
monopole sources related to the moisture of the air. We show that
for the typical parameters of the strong convective storm the
acoustic output of this monopole source is two orders of magnitude
stronger than Lighthill's quadrupole source, whereas the dipole
radiation related to temperature inhomogeneities is of the same
order as radiation of Lighthill's quadrupole source. The dipole
source related to stratification and the dipole and quadrupole
sources related to inhomogeneity of background velocity are
inefficient sources of sound. The total power of the source related
to moisture is of order $10^7~{\rm watts}$ for the typical
parameters of the strong convective storms, in qualitative agreement
with observations \cite{GH,BG00,GG75}.

The paper is organized as follows: In Sec. \ref{sec:2} equations
governing sound generation by turbulence for moist atmosphere are
obtained in the framework of Lighthill's acoustic analogy. Various
sources of acoustic radiation are analyzed in Sec. \ref{sec:3}.
Application of the obtained results to infrasound generation in
strong convective storms is made in Sec. \ref{sec:4}. Conclusions
are given in Sec. \ref{sec:5}.

\section{\label{sec:2} General formalism}

The dynamics of convective motion of moist air is governed by
continuity, Euler, heat, humidity and ideal gas state equations:
\begin{equation}
\frac{D \rho}{D t} + \rho {\bf \nabla} \cdot {\bf
v}=0,\label{eq:21}
\end{equation}
\begin{equation}
\rho\frac{D {\bf v}}{D t} + 2 \rho {\bf \Omega}\times {\bf
v}=-{\bf \nabla}p- \rho {\bf \nabla}\Phi,\label{eq:22}
\end{equation}
\begin{equation}
T\frac{D s}{D t}=-L_\nu \frac{D q}{D t},\label{eq:23}
\end{equation}
\begin{equation}
\rho=\frac{p}{R T}\frac{1}{1-q+q/\epsilon}=\frac{p}{R
T}\frac{1}{1+aq},\label{eq:25}
\end{equation}
where ${\bf v},~2{\bf \Omega}\times {\bf v},~\rho$ and $p$ are
velocity, Coriolis acceleration, density and pressure respectively;
 $D/Dt\equiv\partial/\partial t+{\bf v}\cdot
{\bf \nabla}$ is Lagrangian time derivative; $L_\nu$ is the latent
heat of condensation and $q$ is the mass mixing ratio of water vapor
(humidity mixing ratio)
\begin{equation}
q\equiv \frac{\rho_\nu}{\rho},\label{eq:26}
\end{equation}
where $\rho_\nu$ is the mass of water vapor in the unit volume;
$\epsilon \equiv m_\nu/m_d \approx 0.622$ is the ratio of
molecular masses of water and air; $a=0.608$ and $R$ is the
universal gas constant.

In the set of Eqs. (\ref{eq:21})-(\ref{eq:25}) diffusion and
viscosity effects are neglected due to the fact that they have minor
influence on low frequency acoustic wave generation as well as its
propagation.

In the future analysis we also assume $\Omega = 0$. As it is well
known \cite{B}, Coriolis effects are negligible for mesoscale
convective system dynamics. On the other hand, when the frequency of
acoustic waves $\Omega_a$ satisfy the condition $\Omega_a \gg
\Omega$, Coriolis effects have also negligible influence on acoustic
wave dynamics.

The main idea of Lighthill's acoustic analogy is reformulation of
the governing equations in the form suitable for the study of
acoustic wave radiation process. To proceed in this direction one
have to choose appropriate "acoustic variable", that describes
acoustic waves in the irrotational regions of the fluid. Generalized
Bernoulli's theorem \cite{BE} suggests that the total enthalpy
\begin{equation}
B\equiv E+\frac{p}{\rho} +\frac{v^2}{2}+\Phi,\label{eq:27}
\end{equation}
where $\Phi$ is gravitational potential energy per unit mass, $E$ is
internal energy and ${\bf \nabla}\Phi\equiv-{\bf g}$, is one of the
possible appropriate choices \cite{H01}. $B$ is constant in the
steady irrotational flow and at large distances from acoustic
sources perturbations of $B$ represent acoustic waves.

For derivation of acoustic analogy equation in terms of the total
enthalpy it is useful to rewrite Euler's equation in the Crocco's
form
\begin{equation}
\rho\frac{D {\bf v}}{D t} + {\bf \nabla}B=-{\bf \omega}\times {\bf
v}+ T{\bf \nabla}s,\label{eq:28}
\end{equation}
where ${\bf \omega}$ is vorticity, $T$ is temperature, $s$ is
specific entropy and
\begin{equation}
Tds=dE+pd\left( \frac{1}{\rho} \right) =
dB-\frac{dp}{\rho}-d\Phi-d\left( \frac{v^2}{2}\right).
\label{eq:29}
\end{equation}

From the thermodynamic identity
\begin{equation}
d\rho = \left( \frac{\partial \rho}{\partial p} \right)_{s,q}dp +
\left( \frac{\partial \rho}{\partial s} \right)_{p,q} ds+ \left(
\frac{\partial \rho}{\partial q} \right)_{s,p}dq,\label{eq:210}
\end{equation}
where the subscripts serve as the reminders of the variables held
constant, using Eqs. (\ref{eq:25}) we obtain
\begin{equation}
d\rho = \frac{1}{c_s^2} dp - \frac{\rho}{c_p} ds+
\frac{a\rho}{1+aq} dq, \label{eq:211}
\end{equation}
where
\begin{equation}
c_s \equiv \left( \frac{\partial p}{\partial \rho}
\right)_{s,q}^{1/2}, \label{eq:212}
\end{equation}
is the sound velocity and
\begin{equation}
c_p \equiv T \left( \frac{\partial s}{\partial T} \right)_{p,q}
\label{eq:213}
\end{equation}
is the specific heat of the air.

Eliminating convective derivative of the density from Eq.
(\ref{eq:21}) using Eq. (\ref{eq:211}) we have
\begin{equation}
\frac{1}{\rho c_s^2} \frac{Dp}{Dt} + {\bf \nabla} \cdot {\bf v} =
\frac{1}{c_p} \frac{Ds}{Dt} +\frac{1}{1+aq} \frac{Dq}{Dt}.
\label{eq:214}
\end{equation}
Subtracting the divergence of Eq. (\ref{eq:28}) from time
derivative of Eq. (\ref{eq:214}) and using Eq (\ref{eq:23}) after
long but straightforward calculations we obtain
\begin{eqnarray}
&& \left( \frac{D}{Dt} \left(\frac{1}{c_s^2} \frac{D}{Dt} \right)
-\frac{{\bf \nabla}p \cdot {\bf \nabla}}{\rho c_s^2} -{\bf
\nabla}^2 \right)B= \nonumber\\
&& S_L + S_T + S_q + S_m + S_\gamma, \label{eq:215}
\end{eqnarray}
where $\gamma \equiv c_p/c_v$ is the ratio of specific heats and
\begin{equation}
S_L \equiv \left( {\bf \nabla} + \frac{{\bf \nabla}p}{\rho c_s^2}
\right) \cdot \left( {\bf \omega} \times {\bf v} \right),
\label{eq:216}
\end{equation}
\begin{equation}
S_T \equiv -\left( {\bf \nabla} + \frac{{\bf \nabla}p}{\rho c_s^2}
\right) \cdot \left( T{\bf \nabla}s \right), \label{eq:217}
\end{equation}
\begin{equation}
S_q \equiv \frac{\partial }{\partial t} \left( \frac{\gamma
T}{c_s^2} \frac{Ds}{Dt} \right) + \left( {\bf v} \cdot {\bf
\nabla}\right) \left( \frac{T}{c_s^2} \frac{Ds}{Dt} \right),
\label{eq:218}
\end{equation}
\begin{equation}
S_m \equiv \frac{\partial }{\partial t} \left( \frac{a}{1+aq}
\frac{Dq}{Dt} \right), \label{eq:219}
\end{equation}
\begin{equation}
S_\gamma \equiv p\frac{\partial \gamma}{\partial q} \left(
\frac{\partial q}{\partial t} \left({\bf v} {\bf \nabla} \right) p
- \frac{\partial p}{\partial t} \left({\bf v} {\bf \nabla} \right)
q \right). \label{eq:220}
\end{equation}

Eq. (\ref{eq:215}) is suitable for identification of different
acoustic sources and the study of their acoustic output.

The nonlinear wave operator on the left of Eq. (\ref{eq:215}) is
identical with that governing the propagation of sound in
irrotational, homentropic flow. Therefore the terms on the right may
be identified as acoustic sources. Propagation of infrasound in the
atmosphere was intensively studied by different authors (see, e.g.,
Ref. \cite{OGCG04} and references therein) and will not be
considered in this paper.

For simplification of further analysis of the acoustic output
of different sources we make several standard assumptions:

(a) Studding acoustic wave generation process for low Mach number
flow, all the convective derivatives in Eq. (\ref{eq:215}) can be
replaced by time derivatives $\partial /
\partial t$ \cite{G};

(b) For acoustic waves with the wavelength $\lambda$ not exceeding
the stratification length scale
\begin{equation}
\lambda \lesssim H \equiv \frac{c_s^2}{g} \approx 10^4 {\rm m},
\label{eq:31}
\end{equation}
one can also neglect the influence of stratification on the acoustic
wave generation process and consider background thermodynamic
parameters in Eq. (\ref{eq:215}) as constants \cite{S67}.

(c) Neglecting nonlinear effects of acoustic wave propagation and
scattering of sound by vorticity and taking into account that $M\ll
1$, for the acoustic pressure in the far field we have
\begin{equation}
p^\prime({\bf x},t) \approx \rho_0 B({\bf x},t). \label{eq:32}
\end{equation}

(d) Eq. (\ref{eq:215}) is equivalent to initial set of Eqs.
(\ref{eq:21})-(\ref{eq:25}) and therefore it describes not only
acoustic waves, but also the instability wave solutions that are
usually associated with large scale turbulent structures and
continuous spectrum solutions related to "fine-grained" turbulent
motions \cite{G02,Go84}. In the presence of any kind of
inhomogeneity, such as stratification or velocity shear, linear
coupling between these perturbations is possible, and in principle
acoustic waves can be generated by both instability waves and
continuous spectrum perturbations. But in the case of low Mach
number ($M \ll 1$) flows both kind of perturbations are very
inefficient sources of sound. The acoustic power is proportional to
$e^{-1/2M^2}$ and $e^{-\pi \delta/2 M}$ for instability waves and
continuous spectrum perturbations respectively \cite{CH90}. In
the last expression $\delta$ is the ratio of length scales of energy
containing vortices and background velocity inhomogeneity
($V/\partial_zV$). In the case of supercell thunderstorm $M\sim
0.1-0.15$ and $\delta \sim 10^{-2}$, therefore both linear
mechanisms have negligible acoustic output and attention should be
payed to sources of sound related to nonlinear terms and entropy
fluctuations that will be studied below.

With these assumptions Eq. (\ref{eq:215}) simplifies and reduces to
\begin{eqnarray}
\frac{1}{\rho_0} \left( \frac{1}{c_s^2} \frac{\partial^2}{\partial
t^2} - {\bf \nabla}^2 \right) p^\prime = S_L + S_T + S_\gamma + S_q
+ S_m, \label{eq:33}
\end{eqnarray}
with
\begin{equation}
S_L \approx {\bf \nabla} \cdot \left( {\bf \omega} \times {\bf v}
\right), \label{eq:34}
\end{equation}
\begin{equation}
S_T \approx -{\bf \nabla} \cdot \left( T{\bf \nabla}s \right),
\label{eq:35}
\end{equation}
\begin{equation}
S_\gamma = p\frac{\partial \gamma}{\partial q} \left( \frac{\partial
q}{\partial t} \left({\bf v} {\bf \nabla} \right) p - \frac{\partial
p}{\partial t} \left({\bf v} {\bf \nabla} \right) q \right).
\label{eq:38}
\end{equation}
\begin{equation}
S_m \approx \frac{a}{1+aq} \frac{\partial^2q}{\partial t^2},
\label{eq:37}
\end{equation}
\begin{equation}
S_q \approx -\frac{\gamma L_\nu}{c_s^2} \frac{\partial^2q}{\partial
t^2}, \label{eq:36}
\end{equation}

First three terms on the right hand side of Eq. (\ref{eq:33}) represent
well known sources of sound: the first term represents Lighthill's
quadrupole source \cite{L52}; the second term is dipole source related to
temperature fluctuations \cite{G}; $S_\gamma$ is monopole source
related to variability of adiabatic index, that usually have
negligible acoustic output \cite{H01} and will not be considered in
the presented paper; Eq. (\ref{eq:33}) shows that in the case of
saturated moist air turbulence there exist two additional sources of
sound. $S_q$ and $S_m$ are monopole sources related to nonstationary
heat and mass production during the condensation of moisture,
respectively.

\section{\label{sec:3} Analysis of different sources}

For estimation of different acoustic sources we follow the standard
\cite{G,H01} procedure. Namely, using the wave equation free space
Green function
\begin{equation}
G(t,t^\prime,{\bf x},{\bf x^\prime})=\frac{\delta(t-t^\prime-|{\bf
x}-{\bf x^\prime}|/c_s)}{4\pi c_s^2|{\bf x}-{\bf
x^\prime}|},\label{eq:41}
\end{equation}
acoustic pressure fluctuations corresponding to a source $S_i$ can
be written as
\begin{equation}
p^\prime_i(x,t) =\frac{1}{4\pi c_s^2} \int
\frac{[S_i]_{t=t_\ast} }{|{\bf x}-{\bf x^\prime}|}d^3{\bf
x^\prime},\label{eq:42}
\end{equation}
where $t_\ast=t-|{\bf x}-{\bf x^\prime}|/c_s$.

Calculating acoustic radiation in the far field ($|{\bf x}| \gg
|{\bf x^\prime}|$), we can use following expansions
\begin{equation}
|{\bf x}-{\bf x^\prime}| \approx |{\bf x}| - \frac{{\bf x}\cdot{\bf
x^\prime}}{|{\bf x}|} ,\label{eq:43}
\end{equation}
\begin{equation}
S_\alpha(t_\ast) \approx S_\alpha \left( t-\frac{|{\bf x}|}{c_s}
\right) + \frac{{\bf x}\cdot{\bf x^\prime}}{c_s|{\bf x}|}
\frac{\partial}{\partial t}S_\alpha \left( t-\frac{|{\bf x}|}{c_s}
\right) ,\label{eq:44}
\end{equation}
and using plane wave approximation for far field derivatives
\begin{equation}
\frac{\partial}{\partial x_i} \approx -\frac{x_i}{c_s|{\bf x}|}
\frac{\partial}{\partial t},\label{eq:45}
\end{equation}
for the Lighthill's source we obtain
\begin{equation}
p^\prime_L(x,t) = - \frac{\rho_0x_ix_j}{4\pi c_s^2|{\bf x}|^3}
\frac{\partial^2}{\partial t^2} \int v_iv_j d^3{\bf
x^\prime},\label{eq:46}
\end{equation}
which corresponds to quadrupolar radiation field.

$p_L^\prime$ can be estimated in terms of characteristic velocity
$v$ and length scale $l$ of energy containing turbulent eddies.
Fluctuations in $v_iv_j$ in different regions of the turbulent flow
separated by distances greater then $l$ tend to be statistically
independent, and therefore generation of sound can be considered as
by a collection of $F/l^3$ independent eddies, where $F$ is the
volume occupied by the turbulence. The dominant frequency of the
motion $\sim v/l$, so the wavelength of the radiated sound $\Omega
\sim l/M_t$, where $M_t\equiv v/c_s\ll 1$ is turbulent Mach number.
Therefore, each eddy is acoustically compact. Acoustic pressure
generated by single eddy is $p^\prime_{L1}\sim (l/|{\bf
x}|)\rho_0v^2M_t^2$, and acoustic power it radiates $N_{L1} \sim
4\pi |{\bf x}|^2 p^{\prime 2}_{L1}/\rho_0c_s\approx \rho_0
v^3l^2M^5$, that corresponds to Lighthill's eighth power law. For
total acoustic power this yields Praudman's estimate \cite{P52}
\begin{equation}
N_L \sim \frac{\rho  v^8}{lc_s^5}F. \label{eq:47}
\end{equation}

Similar arguments can be used for estimation of acoustic power of
thermo-acoustical source $S_T$ related to density (and therefore
temperature) fluctuations, that produce dipole source \cite{H01}.
The physics of this kind of acoustic radiation is the following:
"hot spots" or "entropy inhomogeneities" behave as scattering
centers at which dynamic pressure fluctuations are converted
directly into sound. The acoustic power is
\begin{equation}
N_T \sim \frac{\rho \Delta T^2 v^6}{l T^2 c_s^3}F=\frac{\Delta
T^2}{M_t^2 T^2}N_L, \label{eq:48}
\end{equation}
where $\Delta T$ denotes the rms of temperature fluctuations.

Acoustic sources $S_q$ and $S_m$ are related to the moisture of the
air. They produce monopole radiation and physically have the
following nature: suppose there exist two saturated air parcels of
unit mass with different temperatures $T_1$ and $T_2$ and water
masses $m_\nu(T_1)$ and $m_\nu(T_2)$. Mixing of these parcels leads
to the condensation of water due to the fact that
\begin{equation}
2m_\nu(T_1/2+T_2/2) <m_\nu(T_1)+m_\nu(T_2). \label{eq:48a}
\end{equation}

Condensation of the water leads to two effects, important for sound
generation: production of heat and decrease of the gas mass. Both of
these effects are known to produce monopole radiation \cite{G,H01}.
Consequently, turbulent mixing of saturated air with different
temperatures will lead not only to the dipole thermo-acoustical
radiation (\ref{eq:48}), but also to the monopole radiation.

According to Eqs. (\ref{eq:26}) and (\ref{eq:48a}) for humidity mixing
ratio fluctuation $q_s^\prime$ we have
\begin{equation}
q_s^\prime = q_s(T+T^\prime)+q_s(T-T^\prime)-2q_s(T).
\end{equation}
In the limit $T^\prime /T\ll 1$ this yields
\begin{equation}
q_s^\prime \approx \frac{\partial^2 q_s}{\partial T^2}{T^\prime}^2.
\label{eq:412}
\end{equation}
Substituting (\ref{eq:218}) into (\ref{eq:42}) and using
(\ref{eq:412}), (\ref{eq:43}) and (\ref{eq:44}) we obtain
\begin{equation}
p^\prime_q({\bf x},t) = - \frac{\rho_0\gamma L_\nu}{4\pi c_s^2|{\bf
x}|} \frac{\partial^2 q_s}{\partial T^2} \frac{\partial^2}{\partial
t^2} \int T^\prime ({\bf x}^\prime , t)T^\prime ({\bf x}^\prime , t)
d^3{\bf x^\prime},\label{eq:46}
\end{equation}
which corresponds to monopole radiation field.

For total acoustic power radiated by monopole source related to the
moisture we have
\begin{eqnarray}
N_q =\frac {4\pi |{\bf x}|^2}{\rho_0 c_s}\langle p^\prime ({\bf
x},t)p^\prime ({\bf x},t)\rangle\sim \\ \nonumber {\frac {\rho_0 {\gamma}^2
L^2_\nu}{c^5_s}}{\left (\frac {\partial^2 q_s}{\partial T^2}\right
)}^2\frac{\partial^4}{\partial t^4}\int d^3{\bf x^\prime}d^3{\bf
x}^{\prime \prime}\langle T^\prime ({\bf x}^\prime ,t)T^\prime ({\bf
x}^\prime ,t)T^\prime ({\bf x}^{\prime \prime} ,t)T^\prime ({\bf
x}^{\prime \prime} ,t)\rangle \label{eq:A1}
\end{eqnarray}
Fluctuations of temperature in different regions of the turbulent flow
separated by distances greater then length scale $l$ of energy
containing eddies are not correlated and therefore the integral
in Eq.(\ref{eq:A1}) can be estimated as $F_1 l^3\Delta T^4$, where $F_1$
is the volume occupied by saturated moist air turbulence. Taking also into
account that the characteristic timescale of the process is turn over
time of energy containing turbulent eddies $l/v$ finally obtain
\begin{equation}
N_q \sim {\frac {\rho_0 {\gamma}^2 L^2_\nu \Delta q^2
M_t^4}{lc_s}}F_1 = \frac{\gamma^2L_\nu^2\Delta q^2}{M_t^4c_s^4}
\frac{F_1}{F} N_L, \label{eq:49}
\end{equation}
where $\Delta q$ is the rms of humidity mixing ratio perturbations.
the acoustic power of the source related to the gas mass production
we obtain
\begin{equation}
N_m \sim \frac{\rho_0 a^2 c_s^3 \Delta q^2 M_t^4}{l}F_1 = \frac{a^2
\Delta q^2}{M_t^4}\frac{F_1}{F}N_L. \label{eq:410}
\end{equation}

\section{\label{sec:4} Application to infrasound generation by strong convective storms}

In this section we apply our findings to study infrasound generation
by strong convective storms. Taking for typical parameters of supercell storms
$v \sim 5~{\rm m/s}$, $\Delta T \sim 3^\circ~{\rm K}$ \cite{GH,B}, $T=270^\circ~{\rm K}$ and
$c_s=330$ m/s and using Eqs. (\ref{eq:47}) and (\ref{eq:48}) we see
that the dipole radiation related to temperature inhomogeneities is
of the same order as radiation of Lighthill's quadrupole source.

Combining Eqs. (\ref{eq:49})-(\ref{eq:410}), using $L_\nu \approx
2.5\times10^6~{\rm m^2/s^2}$ and $\gamma \approx 1.4$ we obtain
\begin{equation}
\frac{N_q}{N_m} \approx \left[ \frac{\gamma L_\nu}{c_s^2} \right]^2
\approx 10^3, \label{eq:410a}
\end{equation}
therefore acoustic power of the source related to the gas mass
production is negligible compared to the radiation related to the
heat production.

Estimation of $\Delta q$ is a bit more difficult. For saturation
specific humidity we use Bolton's formula \cite{B80}
\begin{equation}
q_s \approx \frac{3.8}{p_0} \exp\left( \frac{17.67
T_c}{T_c+243.5} \right), \label{eq:411}
\end{equation}
where $T_c=T-273.15$ is the temperature in degree Celsius and $p_0$
is atmospheric pressure in ${\rm mb}$. Taking into account (\ref{eq:412})
and using $p_0\approx 800~{\rm mb}$, we obtain
\begin{equation}
\Delta q \approx \frac{6.8 \cdot 10^4}{(243.5+T_c)^4} \exp \left(
\frac{17.67 T_c}{T_c+243.5} \right) \Delta T^2 \equiv f(T_c)
\frac{\Delta T^2}{T^2}. \label{eq:413}
\end{equation}
Note, that due to the numerator in the exponent $f(T_c)$ strongly
depends on temperature, e.g., $f(T_C=10^\circ)/f(T_C=0^\circ)\approx
2$.

Using Eqs. (\ref{eq:47}) and (\ref{eq:49}) we obtain
\begin{equation}
\frac{N_q}{N_L} \approx \left[ \frac{\gamma L_\nu}{c_s^2} \right]^2
\left[ \frac{\Delta T}{M_t T} \right]^4  \left[ \frac{F_1}{F}
\right]^2 f^2(T_c). \label{eq:414}
\end{equation}
For our analysis we assume $T_c=0^\circ {\rm C}$, $(f(0)\approx 1.66)$
and $F\approx 125~{\rm km^3}$ \cite{GG75}. For estimation of $F_1$
we note that that for atmospheric convection saturation
level is usually at the height $\approx 1-1.5~{\rm km}$.
Taking also into account that $f(T_c)$ rapidly drops with the
decrease of $T_c$, one can expect that main acoustic radiation will
be produced at the heights $(1.5-4)~{\rm km}$, and consequently we assume $F_1
\approx 0.5 F$. Then Eq. (\ref{eq:414}) yields
\begin{equation}
\frac{N_q}{N_L} \approx 2\times10^2. \label{eq:415}
\end{equation}
Therefore, we conclude that infrasound radiation of supercell storm
should be dominated by the monopole source related to the heat
production during water condensation. Assuming additionally constant
of proportionality in Eq. (\ref{eq:47}) equal to $100$ \cite{G,GG75}
and $l\approx 10~{\rm m}$ for total power of the radiation we obtain
\begin{equation}
N_q \approx 2.4\times10^7~{\rm watts}, \label{eq:416}
\end{equation}
in qualitative agreement with observations \cite{GH,GG75,BG00}.

As it was mentioned above, the characteristic frequency of the
emitted acoustic waves $\Omega \sim v/l$, and using the
characteristic values of the velocity and length scale we obtain for
the period $\tau \sim 10~{\rm s}$.

\section{\label{sec:5} conclusions}

In the presented paper we have considered acoustic radiation from
turbulent convection in the framework of generalized acoustic
analogy taking into account effects of stratification, temperature
fluctuations and moisture of the air. Analysis shows existence of
monopole sources related to heat and mass production during the
condensation of moisture in the saturated moist air turbulence, in
addition to the Lighthill's quadrupole and known dipole sources of
sound related to stratification and temperature fluctuations. It has
been shown that for the typical parameters of the strong convective
storms infrasound radiation should be dominated by monopole source
related to the moisture of the air. The total power of the source
related to moisture is of order $10^7~{\rm watts}$, in qualitative
agreement with observations of strong convective storms
\cite{GH,BG00,GG75}.

\begin{acknowledgments}

G. G. acknowledges the hospitality of Abdus Salam International Center for Theoretical
Physics. G. G. acknowledges partial support from INTAS
061000017-9258 and Georgian NSF ST06/4-096 and 07/406/4-300 grants.

\end{acknowledgments}

% \thebibliography{}

\end{document}